# Biogeography-Based Optimization of Fuzzy Controllers for Improved Quarter Car Suspension Performance

Lida Shahbandari [1*], Mohammad Mansouri [2]

[1] Department of Computer Engineering, North Tehran Branch, Islamic Azad University, Tehran, Iran

[2] Electrical and Electronic Engineering Department, Shahed University, Tehran, Iran

[1] lida.shahbandari@gmail.com

**Abstract:** This study proposes optimized Type-I and Type-II fuzzy controllers for automotive suspension systems to enhance ride comfort and stability under road disturbances (step/sine inputs), addressing the lack of systematic performance comparisons in existing literature. We integrate Biogeography-Based Optimization (BBO), Particle Swarm Optimization (PSO), and Genetic Algorithms (GA) to tune controller parameters for a quarter car model, with emphasis on BBO's underexplored efficacy. MATLAB Simulink simulations demonstrate that BBO-optimized Type-II fuzzy control reduces body displacement by 22% and acceleration by 18% versus baseline methods under step disturbances, while maintaining computational efficiency. The framework provides practical, high-performance solutions for modern vehicles, particularly electric and autonomous platforms where vibration attenuation and energy efficiency are critical.

## I. INTRODUCTION

The design of automotive suspension systems plays a pivotal role in ensuring passenger comfort, vehicle stability, and safety, making it a cornerstone of modern vehicle engineering. As vehicles navigate diverse road conditions, they encounter disturbances such as potholes, speed bumps, and uneven surfaces, which induce vibrations that affect both passengers and the vehicle's structural integrity. These vibrations, if not adequately mitigated, can compromise ride quality, accelerate component wear, and reduce safety. Suspension systems, classified as passive, semi-active, or active, are critical for attenuating these disturbances, with each type offering distinct trade-offs in performance, cost, and complexity. The interdisciplinary significance of suspension system design spans mechanical engineering, control systems, and human factors, impacting automotive industries, transportation safety, and passenger experience. Recent advancements in intelligent control strategies have opened new avenues for enhancing suspension performance, aligning with the growing demand for smarter, more efficient vehicles in the era of autonomous driving and sustainable transportation [1].

A key challenge in suspension system design is achieving an optimal balance between ride comfort and vehicle stability under varying road conditions [2]. Passive suspension systems, consisting of springs and dampers without feedback control, are cost-effective but limited in adaptability to dynamic disturbances. Active suspension systems, while capable of providing superior vibration control across a wide bandwidth, require significant power and complex hardware, making them expensive and less practical for widespread adoption. Semi-active systems offer a compromise, combining passive components with adaptive control to improve performance without the high costs of fully active systems [3]. However, the complexity of modeling and controlling the nonlinear dynamics of a full vehicle suspension system poses significant challenges. To address this, researchers often employ simplified models, such as the quarter car model, which captures the essential dynamics of a single wheel and its interaction with the vehicle body and road. Despite its simplicity, the quarter car model remains a powerful tool for evaluating control strategies, yet its performance under diverse road inputs and uncertainties remains an open research problem. Techniques for expanding the domain of attraction in nonlinear systems [4] can also be applied to improve control system performance under conditions involving model uncertainties and sensor noise [5].

[6] This study applies an ant colony optimization algorithm alongside nonlinear pushover analysis to enhance the seismic performance of 2D steel chevron-braced frames, achieving greater energy dissipation and load-bearing capacity than conventional designs. Ahmadi et al. provide a comprehensive review of recent progress in unsupervised time-series analysis using autoencoders and vision transformers, highlighting core architectures and a wide range of practical applications [7]. A sensor-based simulation method is proposed for predicting global lateral displacement in reinforced concrete bridge columns during seismic events, incorporating both flexural behavior and bar–slip effects

[8]. Another study introduces a smart infrastructure monitoring system using IoT-enabled dashcams and digital twin technology for real-time detection and reporting of damaged roadside assets, contributing to safer and more responsive maintenance practices [9]. [10] Utilizing the Artificial Bee Colony (ABC) algorithm integrated with performance-based pushover analysis, this research optimizes the seismic design of 2D reinforced concrete wall-frames, aiming for structural weight reduction while accounting for nonlinearity and reinforcement configuration. A probabilistic control framework is introduced using a sequential ensemble Kalman smoother to optimally regulate convolutional neural networks in high-dimensional dynamical systems [11]. A machine learning approach based on eye-tracking data is employed to detect cognitive load in immersive virtual reality training scenarios, enabling the analysis of user engagement and mental effort [12]. A hierarchical Type-II fuzzy controller, fine-tuned through the imperialist competitive algorithm, is proposed for load frequency control, demonstrating improved resilience and dynamic performance in power systems [13]. A quadcopter control scheme inspired by Brain Emotional Learning is developed, showing superior adaptability and performance compared to conventional methods under varying conditions [14]. Lastly, a cellular teaching-learning-based optimization strategy is introduced for solving dynamic multi-objective problems, outperforming current leading algorithms in rapidly changing environments [15].

Recent literature has explored various control strategies to enhance suspension system performance using the quarter car model. Classical approaches, such as proportional-integral-derivative (PID) control, have been widely applied due to their simplicity and effectiveness in linear systems[3]. However, the nonlinear and uncertain nature of vehicle dynamics has driven interest in advanced control techniques, including fuzzy logic, sliding mode control, and adaptive control [16]. Fuzzy logic controllers, in particular, have gained attention for their ability to handle nonlinearities and uncertainties without requiring precise mathematical models. Type-I fuzzy controllers have been successfully implemented to improve ride comfort [17], while Type-II fuzzy systems, which account for uncertainty in membership functions, offer enhanced robustness [18]. Optimization algorithms, such as Particle Swarm Optimization (PSO), Genetic Algorithms (GA), and Biogeography-Based Optimization (BBO), have been employed to tune controller parameters, improving performance metrics like body displacement and acceleration [19]. These studies highlight the potential of intelligent control to address the limitations of traditional methods, yet challenges remain in optimizing controller performance under diverse road conditions and computational constraints.

Despite these advancements, a significant gap persists in the systematic comparison of Type-I and Type-II fuzzy controllers for quarter car suspension systems, particularly when optimized using bio-inspired algorithms. Most studies focus on individual control strategies or optimization techniques without comprehensively evaluating their combined impact on system performance across varied road inputs, such as step and sine disturbances. Furthermore, while BBO has shown promise in other optimization tasks [20], its application to fuzzy controller tuning for suspension systems remains underexplored compared to PSO and GA. This gap limits the understanding of how these algorithms can enhance controller robustness and efficiency, particularly for Type-II fuzzy systems, which offer greater flexibility in handling uncertainties. Addressing this gap is critical for developing cost-effective, high-performance suspension systems that meet the demands of modern vehicles, including electric and autonomous platforms, where energy efficiency and passenger comfort are paramount[21].

This paper aims to address these challenges by proposing and evaluating Type-I and Type-II fuzzy controllers optimized using BBO, PSO, and GA for a quarter car suspension system. The study seeks to minimize body displacement and acceleration under step and sine road disturbances, enhancing ride comfort and vehicle stability. The contributions of this work are as follows:

- **Novel Controller Design**: Development of Type-I and Type-II fuzzy controllers tailored for the quarter car model, leveraging fuzzy logic's ability to handle nonlinear dynamics.

- **Optimization with Bio-Inspired Algorithms**: Systematic optimization of controller parameters using BBO, PSO, and GA, with a focus on BBO's efficacy in achieving superior performance.

- **Comprehensive Performance Evaluation**: Comparative analysis of controller performance under diverse road inputs, validated through MATLAB Simulink simulations and mean squared error (MSE) metrics.

- **Practical Insights**: Identification of the most effective optimization algorithm and fuzzy controller type for real-world automotive applications, balancing performance and computational efficiency.

These contributions advance the state-of-the-art in intelligent suspension control, offering practical solutions for automotive engineers and researchers. The remainder of this paper is organized as follows: Section II presents the quarter car suspension model and its mathematical formulation. Section III details the design of the Type-I and Type-II fuzzy controllers. Section IV describes the simulation setup and performance evaluation. Section V discusses the results and their implications, followed by conclusions and future research directions in Section VI.

## II. Quarter Car Suspension Model

A good automotive suspension system should have satisfactory road holding ability, while still providing comfort when riding over bumps and holes in the road. When the vehicle is experiencing any road disturbance (i.e. pot holes, cracks, and uneven pavement), the vehicle body should not have large oscillations, and the oscillations should dissipate quickly. Designing an automotive suspension system is an interesting and challenging control problem. When the suspension system is designed, quarter model (one of the four wheels) is used to simplify the problem. Figure 1 shows the model of a standard quarter car system. We will consider mass movements and mass acceleration on the vertical axis ignoring the rotational movement of the vehicle. In this model, x1 is Vertical displacement of sprung mass ,x2 is Vertical displacement of unsprung mass and w is Road excitation. The variables in the model of a quarter car are shown in table 1.

Table 1. Constant Values of mathematical model of quarter car

| Symbol | Description | Value |
|---|---|---|
| m1[kg] | Sprung mass | 466.5 |
| m2[kg] | Unsprang mass | 49.8 |
| k1[N/m] | Spring coefficient of body sprung | 5700 |
| k2[N/m] | Spring constant of unsprung mass | 135000 |
| b1[Ns/m] | Damping constant of sprung mass | 290 |
| b2[Ns/m] | Damping constant of unsprung mass | 1400 |

The Eq.(1-2) of this model shown below that obtained using the Newton's second law for each of the two masses are in motion and Newton's third law of their interaction

$$m_1 \ddot{x}_1 = -(b_1(\dot{x}_1 - \dot{x}_2) + k_1(x_1 - x_2)) \quad (1)$$

$$m_2 \ddot{x}_2 = b_1(\dot{x}_1 - \dot{x}_2) + k_1(x_1 - x_2) - b_2(\dot{x}_2 - \dot{w}) - k_2(x_2 - w) \quad (2)$$

From the equations (1) and (2), the state-space function equations can be defined as follows (3)

$$\begin{bmatrix} \dot{x}_1 \\ \dot{x}_2 \\ x_1 \\ x_2 \end{bmatrix} = \begin{bmatrix} \frac{-b_1}{m_s} & \frac{b_1}{m_s} & \frac{-k_1}{m_s} & \frac{k_1}{m_s} \\ \frac{b_1}{m_u} & \frac{b_1+b_2}{m_u} & \frac{k_1}{m_u} & \frac{-k_1}{m_u} \\ 1 & 0 & 0 & 0 \\ 0 & 1 & 0 & 0 \end{bmatrix} \begin{bmatrix} x_1 \\ x_2 \\ x_1 \\ x_2 \end{bmatrix} + \begin{bmatrix} 0 & 0 \\ \frac{b_2}{m_u} & \frac{k_2}{m_u} \\ 0 & 0 \\ 0 & 0 \end{bmatrix} \begin{bmatrix} \dot{w} \\ w \end{bmatrix} \begin{bmatrix} x_1 \\ x_2 \\ x_1 - x_2 \end{bmatrix} \quad (3)$$

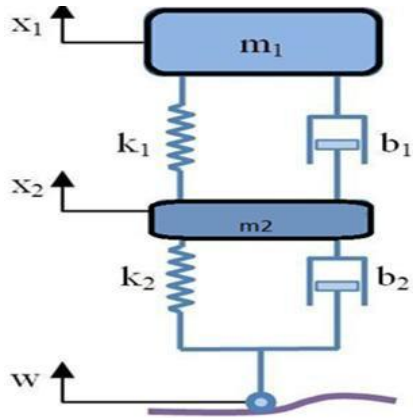

**Figure 1. Passive quarter car suspension system [10]**

### III. Fuzzy Controller Design

The structure of proposed controller has shown in figure 2 .In this paper we use both Fuzzy logic type I and type II controller.

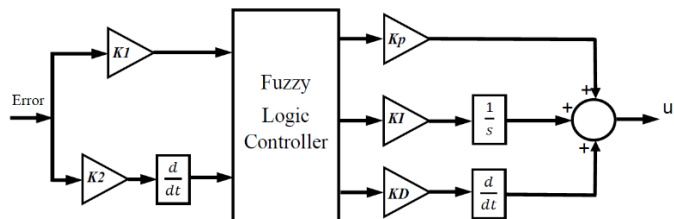

**Figure 2. Structure of fuzzy PID controller for quarter car suspension**

The basic structure of a type-1 fuzzy inference system consists of three conceptual components: a "rule base", which contains a selection of fuzzy rules; a "data base", which defines the membership functions used in the fuzzy rules; and a "reasoning mechanism", which performs the inference procedure upon the rules and given facts to derive a reasonable output or conclusion [11].The List of rules base of the proposed system has shown in table(2). The figure (3) shows the membership functions for type I.

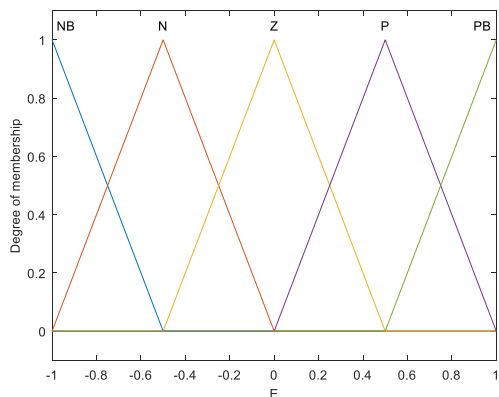

**Figure 3. Membership functions for Inputs in Type I Fuzzy Logic**

The type-2 fuzzy set is a set in which we also have uncertainty about the membership function. Type-2 fuzzy systems consist of fuzzy if-then rules, which contain type-2 fuzzy sets. We can say that type-2 fuzzy logic is a generalization of conventional fuzzy logic (type-1) in the sense that uncertainty is not only limited to the linguistic variables but also is present in the definition of the membership functions [6]. The figure (4) shows the membership functions for type II.

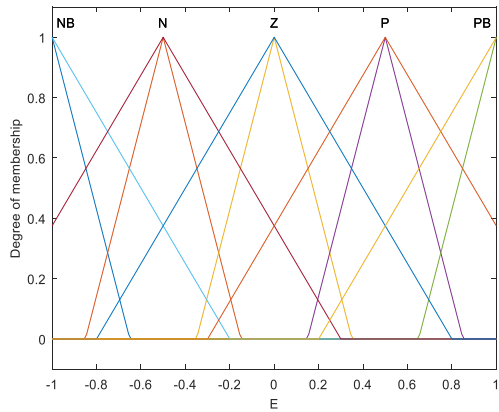

**Figure 4. Membership functions for Inputs in Type II Fuzzy Logic**

We use optimization algorithms like BBO (Biogeography-Based Optimization) , PSO(Particle swarm optimization) and GA(Genetic Algorithms ) to improve the performance of the Fuzzy controller ,though scaling factor and placement of membership functions of Fuzzy set will be optimized .The results shows that using BBO algorithm is more suitable for this issue .By BBO algorithm we optimize the value $\frac{1}{2}$ so that the sum of two coefficients became constant (equal to 1).

**Table 2- Rules of proposed fuzzy logic controller**

| Error | Derivative of Error | Output |
|---|---|---|
| NB | NB | NB |
| NB | N | NB |
| NB | Z | N |
| NB | P | N |
| NB | PB | Z |
| N | NB | NB |
| N | N | N |
| N | Z | N |
| N | P | Z |
| N | PB | P |
| Z | NB | N |
| Z | N | N |
| Z | Z | Z |
| Z | P | P |
| Z | PB | P |
| P | NB | N |
| P | N | Z |
| P | Z | P |
| P | P | P |
| P | PB | PB |
| PB | NB | Z |
| PB | N | P |
| PB | Z | P |
| PB | P | PB |
| PB | PB | PB |

**IV. Simulation**

In this study, the mathematical model as in Eqs. (1-2) of the quarter car suspension system is converted into a computer simulation model in Simulink. Steps and sine inputs have been used to verify system performance. Comparison between the proposed performance of the fuzzy logic controller, which is optimized with BBO and PSO GA, is compared with the proposed controller in [12], as shown in figures 5-8 .

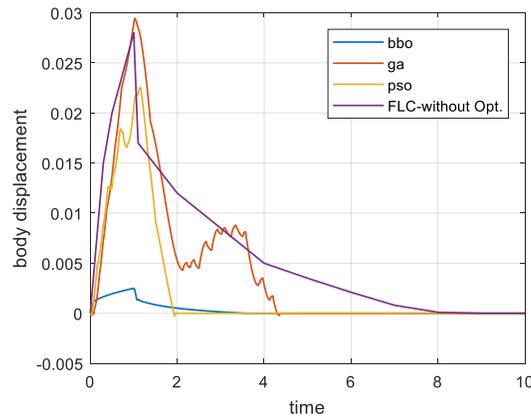

**Figure 5. Body displacement of sprung mass using step unit**

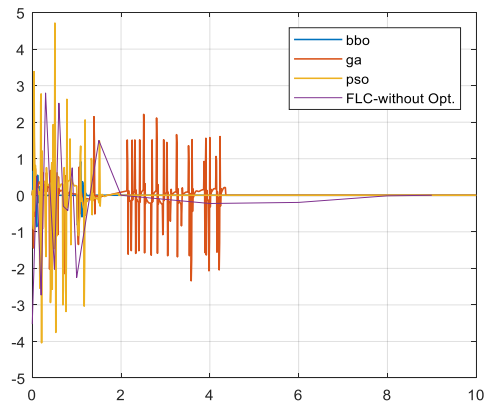

**Figure 6. Body acceleration of sprung mass using step input**

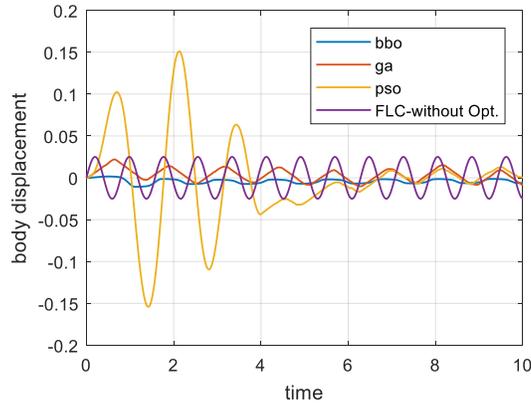

**Figure 7. Body displacement of sprung mass using sine input**

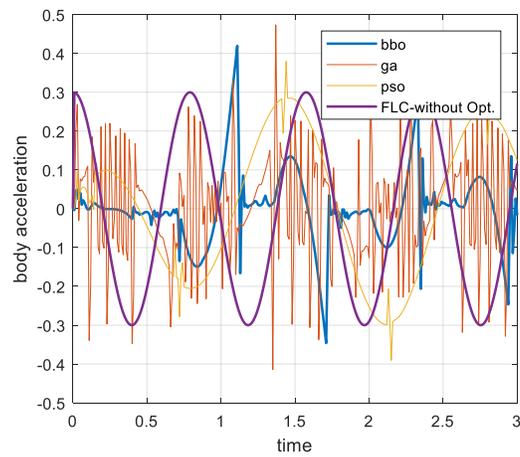

**Figure 8. Body acceleration of sprung mass using sine input**

According to the results, using the BBO algorithm is more suitable for optimizing the controller. Thus, BBO is used to optimize fuzzy logic controller type 2. Result in is shown in figures 9-12.

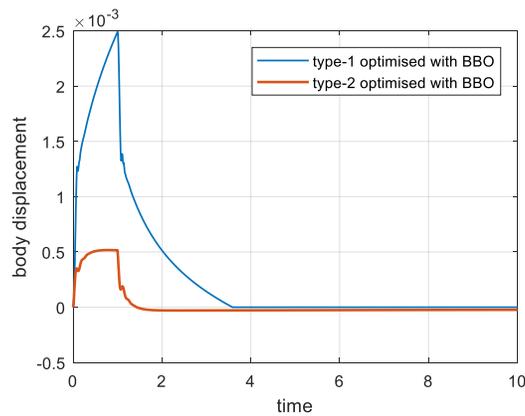

**Figure 9. Body displacement of sprung mass using step unit**

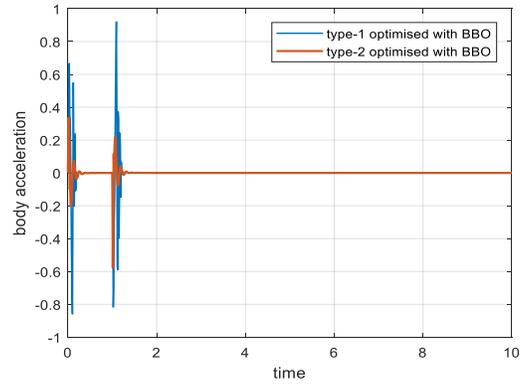

**Figure 10. Body acceleration of sprung mass using step input**

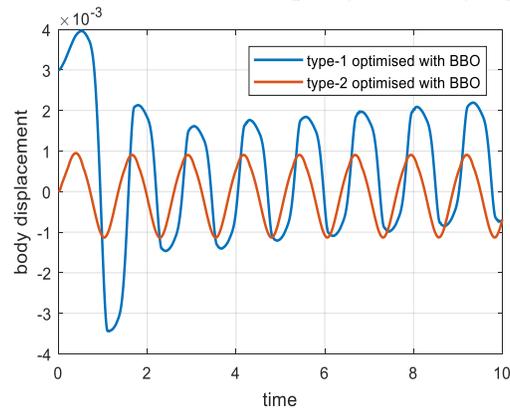

**Figure 11. Body displacement of sprung mass using sine input**

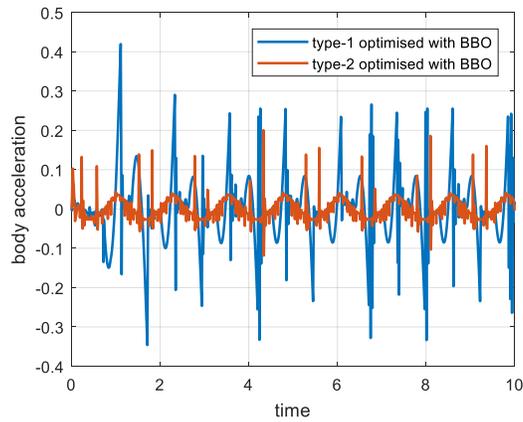

**Figure 12. Body acceleration of sprung mass using sine input**

Also, in order to better compare between the controllers, the MSE criterion is used, the results of which are shown in Table 3.

| Disturbance / controller | Step input | Sine input |
|---|---|---|
| Fuzzy controller proposed in [12] | 21.486 | 12.856 |
| Fuzzy PID type-1 optimised with GA | 37.154 | 25.9313 |
| Fuzzy PID type-1 optimised with PSO | 44.8610 | 15.784 |
| Fuzzy PID type-1 optimised with BBO | 6.7964 | 7.1861 |
| Fuzzy PID type-2 optimised with BBO | 1.5464 | 2.1981 |

## V. Conclusion

According to obtained results from this study, optimization of fuzzy controllers by BBO, PSO and GA algorithms improve the performance of the system and between these three algorithms, BBO algorithm provides better response for system .Also using type II Fuzzy controller cause less movement and acceleration in sprung mass than type I Fuzzy controller to the sine and step disturbance.